\documentclass[prb,twocolumn,showpacs,preprintnumbers,amsmath,amssymb]{revtex4}

\usepackage{color}
\usepackage{graphicx}
\usepackage{dcolumn}
\usepackage{bm}

\preprint{submitted to the PRB}

\begin{document}

\title{Tuning the magnetic properties of half-metallic semi-Heusler alloys by sp-electron substitution:
The case of  AuMnSn$_{1-x}$Sb$_x$ quaternary alloys}

\author{K. \"Ozdo\~gan$^1$}\email{kozdogan@gyte.edu.tr}
\author{I. Galanakis$^2$}\email{galanakis@upatras.gr}
\author{E. \c Sa\c s\i o\~glu$^{3,4}$}\email{e.sasioglu@fz-juelich.de}

\affiliation{$^1$ Department of Physics, Gebze Institute of
Technology, Gebze,
41400, Kocaeli, Turkey\\
$^2$ Department of Materials Science, School of Natural Sciences,
University of Patras,  GR-26504 Patra, Greece \\
$^3$ Institut f\"ur Festk\"orperforschung, Forschungszentrum
J\"ulich, D-52425 J\"ulich, Germany\\
$^4$ Fatih University,  Physics Department, 34500, B\" uy\" uk\c
cekmece,  \.{I}stanbul, Turkey}

\date{\today}

\begin{abstract}
We study the  electronic and magnetic properties of the quaternary
AuMnSn$_{1-x}$Sb$_{x}$ Heusler alloys using first principles
calculations. We determine their magnetic phase diagram and we
show that they present a  phase transition from a ferromagnetic to
an antiferromagnetic state with increasing Sb concentration.  For
large Sb concentrations the antiferromagnetic superexchange
coupling dominates over the ferromagnetic RKKY-like exchange
mechanism. This behavior is similar to the one demonstrated by the
isovalent Ni$_{1-x}$Cu$_x$MnSb alloy studied recently by the
authors [I. Galanakis et al, Phys. Rev. B. \textbf{77}, 214417
(2008)]. Thus the variation of the concentration of the
\textit{sp}-electrons (Sn and Sb atoms) and the variation of the
concentration of the non-magnetic \textit{3d} atoms (Cu) lead to a
similar tuning of the the magnetic properties of the Heusler
alloys. We show that the inclusion of correlation effects does not
alter the phase diagram. Calculated results are in good agreement
with the available experimental data.

\end{abstract}

\pacs{75.50.Cc, 75.30.Et, 71.15.Mb}

\maketitle

Semi-Heusler alloys with the chemical formula XYZ are ternary
compounds consisting of two different transition metals (X and Y)
and one sp element (Z), and crystallize in the C1$_b$ structure.
Their name stems from von Heusler who first studied the Cu-Mn-Al
system in 1903.\cite{Heusler} The interest on these alloys has
been recently revived due the half-metallic character exhibited by
some  semi-Heuslers; i.e. compounds for which one spin channel is
semiconducting or insulating whereas the other has a metallic
character, leading to 100\% carrier spin-polarization at the Fermi
level (E$_F$). Half-metallic ferromagnetism (HMF) was initialy
proposed by de Groot et al. in 1983 when studying the band
structure of the NiMnSb Heusler compound.\cite{de Groot} Also
members of the family of the so-called full-Heusler alloys like
Co$_2$MnZ (Z=Si and Ge) have been predicted to be
half-metals.\cite{Ishida}

Early measurements by Webster et al.  on several quaternary
Heusler alloys, as well as recent studies of Walle et al. on
AuMnSn$_{1-x}$Sb$_{x}$, demonstrated the importance of the
\emph{sp} electrons in establishing the magnetic properties of
Heusler compounds.\cite{LB,AuMnSnSb} On the other hand, the
importance of the non-magnetic 3\textit{d} atoms, like Cu, for the
magnetism of Heusler alloys has been also revealed recently by the
experimental studies of Duong et al. and Ren et al. in the case of
Co$_{1-x}$Cu$_x$MnSb and Ni$_{1-x}$Cu$_x$MnSb alloys,
respectively.\cite{CoCuMnSb,NiCuMnSb}

Authors have shown in Ref. \onlinecite{Galanakis} that when Ni
atoms are substituted by Cu ones in the half-metallic NiMnSb
alloy, first the half-metallicity is lost and for large Cu
concentrations the Ni$_{1-x}$Cu$_{x}$MnSb alloys become
antiferrognetic; the phase transition occurs for a concentration
$x\simeq 0.6$. We have to note here that CuMnSb is a well-known
antiferromagnet and has been extensively studied both
experimentally\cite{Boeuf} and theoretically.\cite{Jeong} The
exchange coupling mechanism  in  Mn-based Heusler alloys has been
well-understood  \cite{SasiogluPT} and  for half-metallic systems
it has been shown in  Ref. \onlinecite{Galanakis} that the
magnetic interactions depend strongly on the position of the Fermi
level within the gap. As the concentration in Cu is increasing the
Fermi level is shifted higher in energy with respect to the
minority-spin gap and at the transition point it crosses enough
the minority-spin conduction band so that the antiferromagnetic
superexchange coupling of the Mn-Mn spin moments through the Cu
atoms dominates over the ferromagnetic RKKY-like interaction
between the Mn atoms.

In the semi-Heusler alloys of the chemical type AuMnZ (Z=In, Sn,
Sb), the magnetization is confined to the Mn sublattice and the Mn
spin moments are well-localized in space due to the large Mn-Mn
distance, i.e. the 3\textit{d} states belonging to different Mn
atoms do not overlap considerably.  M. Amft and P. M. Oppeneer
calculated the largest zero-temperature polar Kerr rotation (-0.45
degree at about 1 eV photon energy) for AuMnSn.\cite{M Amft}
AuMnSn alloy is isovalent (same number of valence electrons in the
primitive unit cell) to NiMnSb and AuMnSb is isovalent to CuMnSb.
Thus the phase magnetic diagram of AuMnSn$_{1-x}$Sb$_{x}$ and
Ni$_{1-x}$Cu$_{x}$MnSb compounds can be directly compared. In this
Brief Report we employ the full--potential nonorthogonal
local--orbital minimum--basis band structure scheme
(FPLO)\cite{fplo} within the local density approximation
(LDA)\cite{perdew} to study the phase diagram of the quaternary
AuMnSn$_{1-x}$Sb$_{x}$ alloys and compare it with our published
results on Ni$_{1-x}$Cu$_{x}$MnSb compounds. We simulate the
disorder within the  the coherent potential approximation (CPA)
framework. Details of the calculations are similar to the ones
presented in Ref. \onlinecite{Galanakis}. We have used the
theoretical calculated equilibrium lattice constants: 5.83\AA\ for
NiMnSb, 5.99 \AA\ for CuMnSb, 6.333 \AA\ for AuMnSn, and 6.464
\AA\ for AuMnSb. These are slightly different than the
experimental ones: 5.93 \AA\ for NiMnSb, 6.09 \AA\ for CuMnSb,
6.341 \AA\ for AuMnSn and 6.379 \AA\ for AuMnSb.\cite{LB,L
Offernes} We should note here that we performed test calculations
also for the experimental lattice constants but results were
almost identical to the case of the theoretical lattice
parameters. We assumed that the lattice constant for the
quaternary alloys varies linearly wit the concentration $x$. We
show that AuMnSn$_{1-x}$Sb$_{x}$ alloys present a similar magnetic
phase diagram with the Ni$_{1-x}$Cu$_{x}$MnSb compounds and the
tuning of the magnetic properties  is insensitive to the origin of
the conduction electrons which mediate the Mn-Mn interactions.

\begin{figure}
\includegraphics[width=\linewidth]{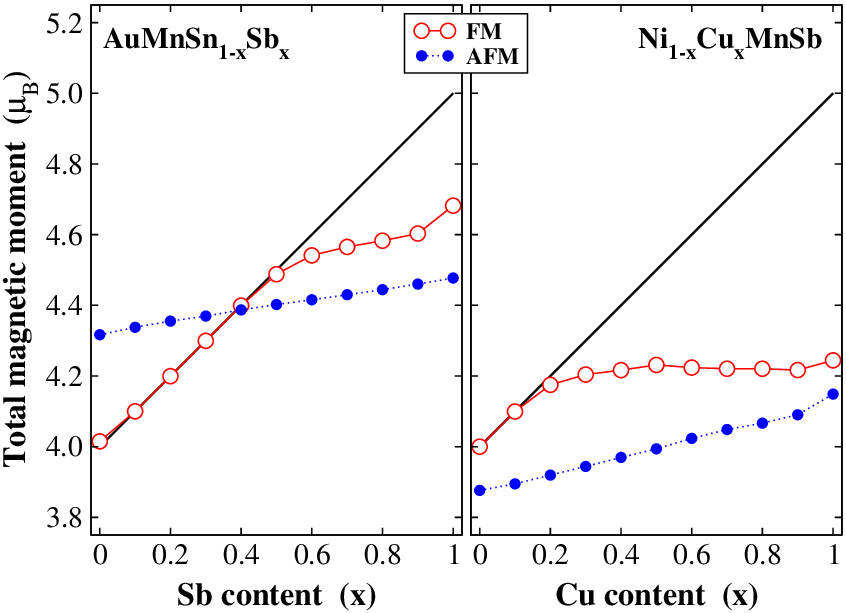}
\caption{ (Color online) Calculated total spin moments (in
$\mu_\mathrm{B}$) as a function of the concentration ($x$) for the
studied AuMnSn$_{1-x}$Sb$_{x}$ and Ni$_{1-x}$Cu$_{x}$MnSb in the
ferromagnetic (FM) configuration. We present also the Mn spin
moment in the antiferromagnetic (AFM) configurations for
comparison.  The solid black lines represent the Slater-Pauling
behavior.} \label{fig1}
\end{figure}

\begin{figure}[t]
\includegraphics[width=\linewidth]{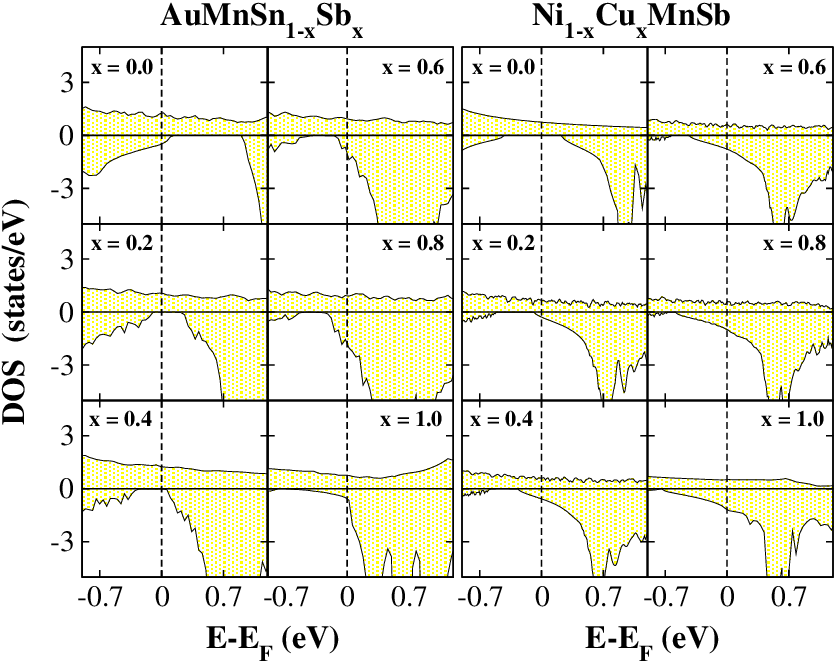}
\caption{(Color online) Spin-resolved total density of states
(DOS) in the case of the AuMnSn$_{1-x}$Sb$_{x}$ (left panel) and
Ni$_{1-x}$Cu$_{x}$MnSb (right panel) around the Fermi level for
selected values of $x$. We have set the Fermi level as the zero of
the energy axis. Positive values of DOS correspond to the
majority-spin electrons and negative values to the minority-spin
electrons.} \label{fig2}
\end{figure}

We will start our discussion from the calculated spin magnetic
moments presented in Fig.~\ref{fig1}. We have drawn the total
magnetic moments for the AuMnSn$_{1-x}$Sb$_{x}$ and
Ni$_{1-x}$Cu$_{x}$MnSb alloys as a function of the concentration
for the ferromagnetic state in Fig.~\ref{fig1}. The solid black
lines represent the Slater-Pauling (SP) behavior obeyed by the
perfect half metallic ferromagnets (the total spin moment in
$\mu_B$ is the number of valence electrons minus 18)
\cite{GalaHalf}. We present the Mn magnetic moment corresponding
to the antiferromagnetic state for comparison. We should note that
the spin magnetic moments of Au, Ni and Cu are zero in the
antiferromagnetic (AFM) state due to symmetry reasons while the Sb
and Sn atoms have a very small magnetic moment value. For $x=0$
the AuMnSn compound has a total spin moment slightly larger than
the ideal 4 $\mu _{B}$ predicted by the SP rule for the perfect
half-metallic ferromagnets since the Fermi level is slightly below
the gap as can be seen in Fig. \ref{fig2} where we have drawn the
total density of state (DOS) for both families of compounds.
NiMnSb is an ideal half-metal and this is reflected on an integer
value of the total spin magnetic moment which is 4 $\mu _{B}$ and
the Fermi level is located in the middle of the gap. The
AuMnSn$_{1-x}$Sb$_{x}$ and Ni$_{1-x}$Cu$_{x}$MnSb follow the SP
rule up to  $x\simeq 0.5$ and $x\simeq 0.2$, respectively, and at
this point the half-metallicity is lost since the Fermi level is
shifted and now crosses the minority-spin conduction band. This is
clearly shown in Fig. \ref{fig2}. The shift of the Fermi level
towards higher energies is easily understood. When we increase the
concentration of Sb and Cu atoms, we dope the system with $p$
charge. The corresponding majority-spin $p$ states are extremely
extended in energy and thus, when their occupation increases, they
push the Fermi level higher in energy.

\begin{table*}[tbp]
\caption{Total and atom-resolved spin magnetic moments in $\mu_B$
for the ferromagnetic configuration as a function of the
concentration, $x$, in AuMnSn$_{1-x}$Sb$_{x}$ and
Ni$_{1-x}$Cu$_{x}$MnSb compounds. Values have been scaled to one
atom.} \label{table1}
\begin{ruledtabular}
 \begin{tabular}{l|ccccc|ccccc}
    & \multicolumn{5}{c|}{AuMnSn$_{1-x}$Sb$_{x}$ (FM)} & \multicolumn{5}{c}{Ni$_{1-x}$Cu$_x$MnSb (FM)}  \\
    $x$ &  Total  &   Au &   Mn &Sn &Sb &  Total  &   Ni &   Cu &Mn
&Sb\\ \hline
  0.0  &  4.015& -0.022& 4.192& -0.155&       & 4.000& 0.255&      & 3.847 & -0.102\\
  0.2  &  4.199&  0.023& 4.286& -0.117& -0.078& 4.175& 0.310& 0.072& 3.982 & -0.070\\
  0.4  &  4.399&  0.064& 4.392& -0.073& -0.033& 4.216& 0.324& 0.078& 4.042 & -0.052\\
  0.6  &  4.541&  0.094& 4.459& -0.038&  0.004& 4.224& 0.327& 0.080& 4.083 & -0.038\\
  0.8  &  4.582&  0.103& 4.470& -0.026&  0.018& 4.220& 0.327& 0.083& 4.113 & -0.025\\
  1.0  &  4.682&  0.118& 4.520&       &  0.044& 4.244&      & 0.096& 4.158 & -0.010\\

\end{tabular}
\end{ruledtabular}
\end{table*}

To reveal the mechanism for the loss of the half-metallic
character we have to study the atom-resolved spin magnetic moments
for the ferromagnetic configuration presented in Table
\ref{table1}. The spin magnetic moment of Au, Sn and Sb atoms
changes from -0.022, -0.155 and -0.100 to 0.118, -0.020 and 0.044
$\mu_B$, respectively, with increasing of the Sb concentration in
the AuMnSn$_{1-x}$Sb$_{x}$ compound. These atoms have almost
filled electronic shells, since they provide electronic bands,
much deeper in energy than the Mn ones,\cite{GalaHalf} and they
contribute marginally to the total spin moment. Thus the extra
electron provided by the Sb atom has to be accommodated by the
bands provided by the Mn atom.  For AuMnSn, the spin magnetic
moment of Mn is 4.192 $\mu_B$ and thus most of the five
majority-spin states are occupied. To occupy further the Mn
majority states costs a lot in energy and the system prefers to
occupy partially also the minority-spin states above the gap and
the half-metallicity is lost. As a result the  Mn spin moment is
only around $0.32 \mu_B$ larger in AuMnSb reaching a value of
4.520 $\mu_B$. The calculated magnetic moments are in good
agreement with experimental values of AuMnSn\cite{A. Neumann} and
AuMnSb\cite{J. Pierre} compounds. In the case of
Ni$_{1-x}$Cu$_{x}$MnSb alloys the same phenomenon occurs: Ni, Cu
and Sb atoms carry very small spin moments and Mn increases its
spin moment by 0.31 $\mu_B$ when all Ni atoms are substituted by
Cu ones leading to the loss of the half-metallicity.

\begin{figure}[t]
\includegraphics[width=\linewidth]{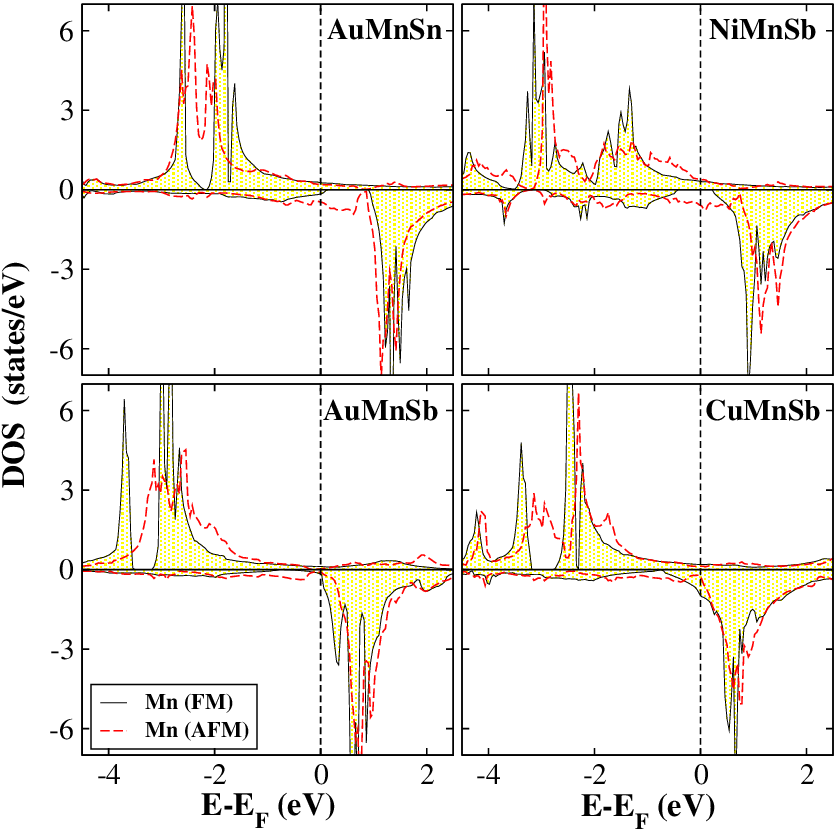}
\caption{(Color online) Spin-resolved Mn density of states (DOS)
of CuMnSb, NiMnSb, AuMnSn and AuMnSb compounds for FM and AFM
configurations. } \label{fig3}
\end{figure}

In the antiferromagnetic (AFM) state, Sb and Sn have a very small
magnetic moment while the magnetic moment of Au is zero in
AuMnSn$_{1-x}$Sb$_{x}$ compound due to symmetry reasons. Similarly
in the case of the Ni$_{1-x}$Cu$_{x}$MnSb alloys, Ni and Cu have
zero spin moments, while Sb has a very small magnetic moment. The
closeness in value between the Mn spin moments in FM and AFM
configurations shown in Fig. \ref{fig1} can be understood if we
examine the Mn-resolved DOS shown for all four CuMnSb, NiMnSb,
AuMnSn and AuMnSb compounds in Fig.~\ref{fig3}. Both in the FM and
AFM cases Mn atoms present a similar DOS and the small broadening
of the bands in the AFM state occurs due to the stronger
hybridization with the other atoms in this case. The important
point is that the similar DOS in the FM and AFM cases justifies
the use of the Anderson $s-d$ model to interpret the results on
the quaternary AuMnSn$_{1-x}$Sb$_{x}$ alloys as in Ref.
\onlinecite{Galanakis} for the  Ni$_{1-x}$Cu$_{x}$MnSb alloys.

\begin{figure}[t]
\includegraphics[width=\linewidth]{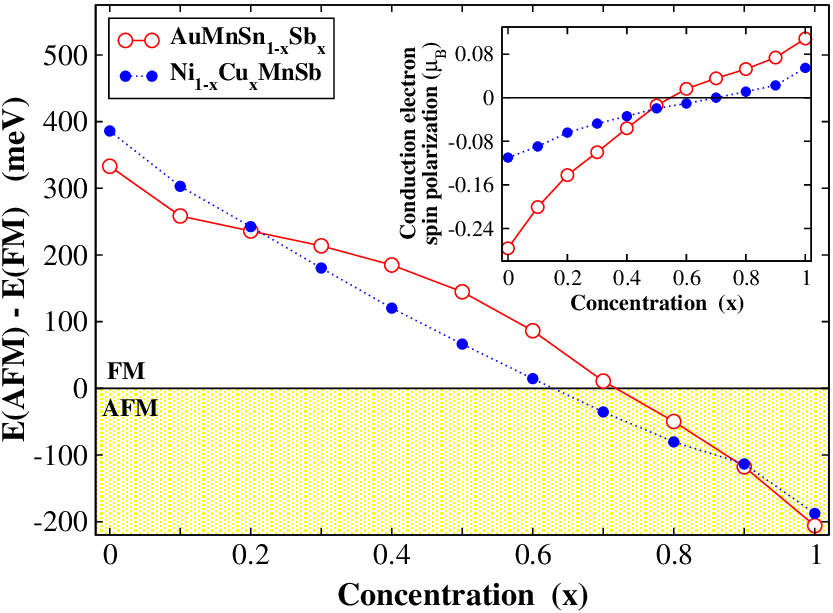}
\caption{ (Color online) Ground state magnetic phase diagram and
total energy differences between AFM and FM configurations of the
Mn magnetic moments in AuMnSn$_{1-x}$Sb$_{x}$  and
Ni$_{1-x}$Cu$_{x}$MnSb as a function of the concentration ($x$).
In the inset we show the total spin polarization of the conduction
electrons of X (Au, Ni, Cu) and Z (Sn and Sb) atoms as a function
of the concentration ($x$). Note that the energy differences are
given for an antiferromagnetic unit cell, while spin polarization
of the conduction electrons is given for a FM unit cell}
\label{fig4}
\end{figure}

As we have  mentioned above both AuMnSn$_{1-x}$Sb$_{x}$ and
Ni$_{1-x}$Cu$_{x}$MnSb quaternary compounds loose their half
metallic character at a concentration of $x\simeq 0.5$ and
$x\simeq 0.2$, respectively. For these values of the concentration
the Fermi level enters the minority-spin conduction band but the
ferromagnetism is still favorable with respect to the AFM state.
To study the phase transition, we have calculated the total
energies for both the FM and the AFM configurations of the Mn spin
magnetic moments.  All energy calculations have been performed
using  the large AFM unit cell (which is double the FM unit cell).
We determine the zero temperature magnetic phase diagram as the
difference of the corresponding total energies ($E_{AFM}-E_{FM}$)
per AFM unit cell and we present our results in Fig.~\ref{fig4}.
AuMnSn$_{1-x}$Sb$_{x}$ shows a phase transition from a FM to an
AFM coupling of the Mn spin moments for a critical concentration
value $x\simeq 0.7$. As seen in Figs.~\ref{fig2} and \ref{fig3}
when we substitute Sb for Sn,  the Fermi level moves towards
higher energies and the number of the minority states just above
the Fermi level increases. This gives rise to an opposite behavior
in the relative contributions of the exchange mechanisms: a
decrease for the RKKY-like coupling and an increase in the
superexchange mechanism. At the transition point both mechanisms
cancel each other and further increase of $x$ leads to an
antiferromagnetic order due to the dominating character of the
superexchange mechanism. The value for the transition for the
AuMnSn$_{1-x}$Sb$_{x}$ alloys is very close to the transition
value of $x\simeq 0.6$ calculated already for
Ni$_{1-x}$Cu$_{x}$MnSb in Ref. \onlinecite{Galanakis} and
reproduced here. The similar behavior of the two families of
alloys can be traced in the spin-polarization of the conduction
electrons of the  Au,Sn and Sb atoms (Ni,Cu,Sb in the case of
Ni$_{1-x}$Cu$_{x}$MnSb) presented in the inset of Fig. \ref{fig4}
which is similar for both families of alloys. These electrons are
responsible for the coupling between the distinct Mn localized
spin magnetic moments. The role of the spin-polarization of the
conduction electrons and their connection to the phase diagram
have been largely discussed in Ref. \onlinecite{Galanakis}.

We should finally note that we have examined also the influence of
the electron-correlation on the magnetic phase diagram for the
AuMnSn$_{1-x}$Sb$_{x}$ alloys taken into account using the popular
LDA+$U$ scheme.\cite{LDA+U} We have assumed a value for the Mn
exchange-splitting constant $J$ of 0.8 eV and have varied the
on-site Coulomb repulsion constant $U$ between 3 and 5 eV and
calculated the $E_{AFM}-E_{FM}$ for AuMnSb. LDA+$U$ should push
the Mn unoccupied minority states higher in energy similarly to
the effect of the expansion of the lattice constant. But for
AuMnSb this shift of the states is very small revealing that
correlations do not play a crucial role for the description of
these alloys. As a result the $E_{AFM}-E_{FM}$ varies between
-156.15 and -155.39 meV as $U$ changes from 3 to 5 eV (close to
the value of $\sim$ -200 meV when $U$ is not included in the
calculations) and the AFM state remains the ground state. Thus the
inclusion of the correlations in our electronic-structure
calculations only slightly shifts the transition point to lower Sb
concentration.

We have studied the effect of the variation of the concentration
of the \textit{sp} electrons on the electronic and magnetic
properties of the quaternary AuMnSn$_{1-x}$Sb$_{x}$ ($0 \leq x
\leq 1$) Heusler alloys using first principles calculations. We
determine their magnetic phase diagram and we show that they
present a phase transition from a ferromagnetic to an
antiferromagnetic state with increasing Sb concentration. For
large Sb concentrations the antiferromagnetic superexchange
coupling dominates over the ferromagnetic RKKY-like exchange
mechanism. Electronic correlation effects have a marginal effect
on the magnetic phase diagram of these compounds. This is an
alternative route for tuning the magnetic properties of the
Heusler alloys with respect to the variation of the non-magnetic
$3d$ atoms shown for Ni$_{1-x}$Cu$_x$MnSb alloys [I. Galanakis et
al, Phys. Rev. B. \textbf{77}, 214417 (2008)]. These findings can
be used as a practical tool to design materials with given
physical properties.

Authors acknowledge the computer support of the Leibniz Institute
for Solid State  and Materials Research Dresden, and the
assistance of Ulrike Nitzsche in using the computer facilities.

\end{document}